# Magnetic-Field-Dependent Raman Scattering in Multiferroic Bilayer Films: Evidence for Stress-Mediated Magnetoelectric Coupling

Zheng Li, Yao Wang, Yuanhua Lin, and C. W. Nan[a]

*Department of Materials Science and Engineering and State Key Laboratory of New Ceramics and Fine Processing, Tsinghua University, Beijing 100084, China*

We report the first magnetic-field-dependent Raman scattering studies on $Pb(Zr,Ti)O_3$-$CoFe_2O_4$ bilayer multiferroic system. The phonon frequencies of the nano-bilayers obviously change with magnetic field, which is absent in the single $Pb(Zr,Ti)O_3$ films. The magnetostriction of the $CoFe_2O_4$ layer generates stress mechanically transferred to the $Pb(Zr,Ti)O_3$ layer, resulting in the mode changes. The observed magnetic-field-induced softening of the soft mode in the $Pb(Zr,Ti)O_3$ layer bears a striking resemblance to direct magnetoelectric output in the bilayers, providing evidence for stress-mediated magnetoelectric coupling mechanism in the multiferroic bi-layers.

PACS numbers: 75.80+q, 77.84.Lf, 63.20.-e, 77.84.Dy

---
[a] Corresponding author: cwnan@tsinghua.edu.cn

Multiferroic magnetoelectric (ME) composite systems with ferro-/ferri-magnetic (e.g., (La,Sr)MnO$_3$, ferrites MFe$_2$O$_4$, M=Co, Ni, etc.) and ferroelectric (e.g., BaTiO$_3$, Pb(Zr,Ti)O$_3$, etc.) compounds combined together, have recently stimulated ever-increasing research activities for their technological applications in novel multifunctional devices such as sensors, transducers and memories [1-4]. The importance of composite multiferroics follows from the fact that none of the existing single phase multiferroic materials [5,6] combine large, robust electric and magnetic polarizations at room temperature but the composites could exhibit a large room-temperature ME effects, i.e., direct ME effect characterized by magnetic control of dielectric polarization or inverse ME effect described by electrical control of magnetism. Among them, ME composite films have become an important topic of ever-increasing interest in last few years [7], since they are easy to be on-chip integration. Potential applications based on such ME layered films have been of particular interest recently. For example, electric voltage control of magnetism demonstrated in a bilayer heterostructure of multiferroic BiFeO$_3$ and magnetic CoFe layers [8] leads to a prototype of ME random access memories [9]. A prototype ME read head was also demonstrated based on the direct ME effect observed in ferroelectric-magnetic bilayers.

It has been widely accepted that ME coupling in bulk ME composites (including the film-on-substrate composites with ferro-magnetic (-electric) films grown on ferro-electric (-magnetic) substrates [11-12]) is strain/stress-mediated [4]. In such structures an applied magnetic (or electric) field induces strain in the magnetic (or ferroelectric) constituent which is mechanically transferred to the ferroelectric (or magnetic) constituent, where it induces a dielectric polarization (or magnetization). By comparison, in the ME layered films, several mechanisms responsible for the ME effect has been proposed. In addition to the strain/stress, the ME effect could occurs by purely electronic origin at interface or surface [14,15], exchange bias [8], and even a mechanism combined magnetoresistance with Maxwell-Wagner interface relaxation [16]. It is understood that substrate-imposed mechanical clamping could suppress the strain-mediated ME coupling [7,17], which makes the strain/stress-mediated mechanism an arguable question. However, experimental work on the ME coupling mechanisms in the layered ferroelectric-magnetic films of technological importance remains quite limited.

In this Letter, we use magnetic field-dependent Raman scattering as a direct probe of the stress-medicated ME coupling in the layered ferroelectric-magnetic oxide films, e.g., bilayer Pb(Zr$_{0.52}$Ti$_{0.48}$)O$_3$ (PZT)-CoFe$_2$O$_4$ (CFO) films. Our investigation reveals obvious magnetic-field-induced changes in phonon frequencies of the bilayer PZT-CFO films, which is precipitated by the mechanically transferred stress/strain from the magnetostrictive CFO layer. The ferroelectric polarization $P$ is proportional to the soft mode displacement, and thus to the soft mode frequency [18,19]. Therefore, this study allows us to directly probe the ME coupling behind via changes in the soft mode frequency induced by magnetic field.

Bilayer PZT-CFO thin films were grown on (111)Pt/Ti/SiO$_2$/Si substrate via a simple sol-gel spin-coating method. The PZT (0.2 M) and CFO (0.1 M) precursor solutions were prepared by the same processing as reported previously [20]. The PZT layer was first coated on the substrate with pyrolyzing and annealing at 400 °C and 700 °C for 5 min, respectively. Then the CFO layer was grown on the PZT layer via the same processing. In the resultant PZT-CFO bilayer films, a bottom PZT layer of ca. 130 nm in thickness exhibits (111) preferential orientation induced by (111)Pt [20], while the top CFO layer consisting of randomly-oriented grains has a thickness of ca. 110 nm, as illustrated by X-Ray diffraction (Rigaku D/max-2500) and electron microscopy (FE-SEM, Hitachi S5500) observation (not presented here). For comparison, the single-phase PZT and CFO films were also grown via the same processing. All the films were grown and maintained under the same condition.

Room-temperature Raman measurements were performed using Micro-Raman spectroscopy (RM2000, Renishaw) in the conventional backscattering and in close-to-90º configurations. An $Ar^+$ laser of 514.5 nm was used as an excitation source. The resolution is about 1 $cm^{-1}$. During the optical measurement, the NdFeB magnets were used and placed at the bottom of the sample glass-holder of the Raman microscope to provide a magnetic bias. The magnetic field bias measured by a Gauss meter (Lakeshore DSP550) was normal to the film plane. That is, an out-of-plane magnetic bias up to 0.28 T was applied on the films. In order to avoid heating of the sample, the power of the incident laser beam was kept below 10 mW. Ferroelectric and ferromagnetic properties were measured by using TF analyzer 2000 (aixACCT Co.) and vibrating sample magnetometer (RikenDenshi, BHV-50HTI), respectively. The direct ME effect was measured as before [21], i.e., electric signals generated from the bilayer films induced by the magnetic field were measured through a lock-in amplifier (SRS. Inc., SR830).

The Raman spectra obtained without magnetic field for the single PZT or CFO films and PZT-CFO bilayer film are shown in Fig. 1. As seen, the Raman spectrum of the PZT-CFO bilayer film is just superposition of two sets of phonon spectra from PZT and CFO. The peaks for CFO can be identified as $E_g$, $T_{1g}$ and $A_{1g}$ modes in cubic CFO spinel [22]. The peaks observed for PZT are mainly attributed to its transverse optical (TO) modes, i.e., $A_1$ and E modes of tetragonal perovskite with *P4mm* space group, which are similar to reported in the literature [18,23]. Although the modes below 100 $cm^{-1}$ were not recorded due to limitation of the equipment and the modes above 250 $cm^{-1}$ for the PZT layer were merged together with the peaks for the CFO layer, other active modes such as the soft-mode $A_1(TO_1)$ and $E(TO_2)$ for the PZT layer, and $A_{1g}$ mode for the CFO layer in the bilayer film are quite clearly distinguished. For PZT, the soft-mode $A_1$(1TO) originates from displacement of the B site atom (Zr or Ti) relative to oxygen octahedral atoms (see Fig. 1b), parallel to the direction of polarization, and has close relation to ferroelectric property [18]. $E(TO_2)$ corresponds to a vibration of the A site atom (Pb) against $BO_6$ octahedra (see Fig. 1b) in the direction perpendicular to polarization. As shown in Fig. 1(b), at $H$=0, in comparison with the single phase PZT or CFO films, the soft-mode $A_1(TO_1)$ frequency of the PZT layer in the bilayer film does not change, while the $E(TO_2)$ phonon of the PZT layer and $A_{1g}$ mode for the CFO layer shift to a bit lower frequencies, mainly arising from different residual stress statuses [24] in the single phase and bilayer films. The same soft-mode frequency in the PZT film and PZT-CFO film implies the PZT layer in the PZT-CFO bilayer exhibits the same ferroelectric polarization as the single PZT film; a bit lower $E(TO_2)$ frequency in the PZT layer indicates [24] that the PZT sandwiched between the substrate and top CFO layer suffers a higher in-plane compressive residual stress than the single PZT film, resulting in a higher appearance coercivity [20], as shown in Fig. 2(a). By comparing with the single CFO film, the top CFO layer in the PZT-CFO film becomes less compressive (i.e., more relaxed) due to the bottom PZT layer also acting as a buffer layer [20], resulting in a lower coercivity (see Fig. 2(b)).

After applied a magnetic field $H$ on the bilayer film, changes in these active modes are observed (Figure 3). First, let us see the modes for the CFO layer. Its $T_{2g}$ and $A_{1g}$ modes, originating from vibrations of the tetrahedral and octahedral sub-lattices in the spinel structure [25], shift to high frequencies with $H$ (Table I), obviously indicating changes in the lattice of the CFO layer upon applying $H$, which is due to its magnetostriction feature. The magnetic domain reorientation involved can also contribute to the wavenumber shift. The magnetoelastic effect of the CFO layer can cause stress at the interface between the two layers. Due to negative magnetostriction of CFO, CFO would shrink a bit along $H$ but expand a bit in the direction normal to $H$. Thus with increasing $H$, the CFO layer suffers increasing in-plane (i.e., normal to $H$) compression.

Of particular interest is that the $A_1$ and E phonons of the PZT layer in the bilayer film also change with $H$, i.e., the soft mode $A_1(TO_1)$ becomes softer, while the $E(TO_2)$ mode becomes harder. By contrast, the $A_1$ and E phonons of the single PZT film do not change with $H$, as shown in Fig. 3(b). For clear illustration, a magnified view of the soft mode $A_1(TO_1)$ is shown in Fig. 3(c). In comparison with $H$-independent Raman spectrum of the single PZT film, the PZT-CFO bilayer exhibits an obvious $H$-dependent Raman scattering. During the Raman measurement, all stress history in the films keeps unchanged except the stress resulting from magnetostriction. The magnetostriction of the top CFO layer would impose an in-plane tensile stress to the bottom PZT layer across the interface. This transferred stress in the PZT layer from the CFO layer is roughly up to about 30 MPa by using elastic moduli and saturation magnetostriction for CFO [17]. A downward or upward frequency-shift in the hard modes has been observed in the PZT films under compressive or tensile stress [24]. The transferred stress from the CFO layer could induce the domain reorientation in the PZT owing to ferroelastic effect and the relaxing of residual stress during domain reorientation, which contributes to the wavenumber shift. Thus the magnetic-field-induced soft-mode softening and $E(TO_2)$-mode hardening are a result of the magnetoelastic effect of the top CFO layer, i.e., an elastic interaction between the top CFO and bottom PZT layers results in these changes in the phonon modes of the PZT layer.

The soft mode $A_1(TO_1)$ is of particular importance, since the polarization $P$ is proportional to the soft mode displacement, and thus to the soft mode frequency, i.e., $P \propto \omega_{A1(TO1)}$ [18]. This qualitative relation leads to the following simplified expression for the $H$-induced polarization variation:

$$[P(0)-P(H)]/P(0) \propto [\omega(0)-\omega(H)]/\omega(0) \tag{1}$$

Fig. 4 shows experimental dependencies of the variations of the $H$-induced $P$ and the $A_1(TO_1)$ mode frequency. The remarkable similarity between them also illustrates that the mechanism of the ME coupling in the PZT-CFO bilayer films is, as in the bulk, due to the elastic interaction between these two layers. Thus the magnetic-field-induced soft-mode softening observed in the bilayer film clearly demonstrates the magnetic-field-induced electric polarization via the elastic interaction, i.e., stress-mediated ME coupling.

In conclusion, we have presented magnetic-field-dependent Raman scattering of the PZT-CFO bilayer films exhibiting good coexistence of ferroelectric and ferromagnetic properties. We have demonstrated that the soft-mode frequency of the PZT layer in the PZT-CFO films is the same as that in the single PZT films, and it decreases with magnetic field, but by contrast the soft-mode frequency in the single PZT films does not change with magnetic field. The $E(TO_2)$ mode in the PZT layer in the PZT-CFO films becomes hardening with magnetic field. The magnetic-field-induced changes in the phonon modes in the PZT-CFO films are due to the magnetostriction of the top CFO layer, which generates stress mechanically transferred to the PZT layer. The magnetic-field-induced softening of the soft mode in the PZT layer is consistent with different ME signal in the films, which provides a direct evidence of stress-induced ME coupling mechanism in the multiferroic bilayer films.

**Acknowledgements**

This work was supported by the NSF of China (Grant No. 50621201 & 50832003) and the National Basic Research Program of China (Grant No. 2009CB623303).

**References**

[1] M. Fiebig, J. Phys. D 38, R123 (2005).

[2] W. Eerenstein, N. D. Mathur and J. F. Scott, Nature 442, 759 (2006).

[3] R. Ramesh and N. A. Spaldin, Nat. Mater. 6, 21 (2007).

[4] C. W. Nan et al., J. Appl. Phys. 103, 031101 (2008).

[5] S. W. Cheong and M. Mostovoy, Nat. Mater. 6, 13 (2007).

[6] P. G. Radaelli et al., Phy. Rev. Lett. 101, 067205 (2008); H. J. Xiang et al., ibid, 101, 037209 (2008); Y. Tokunaga et al., ibid, 101, 097205 (2008).

[7] H. Zheng et al., Science 303, 661 (2004).

[8] Y. H. Chu et al., Nat. Mater. 7, 478 (2008)

[9] M. Bibes and A. Barthelemy, Nat. Mater. 7, 425 (2008)

[10] Y. Zhang et al., Appl. Phys. Lett. 92, 152510 (2008).

[11] W. Eerenstein et al., Nat. Mater. 6, 348 (2007).

[12] C. Thiele et al., Phys. Rev. B 75, 054408 (2007).

[13] J. Wang et al., J. Appl. Phys. 104, 014101 (2008).

[14] J. M. Rondinelli, et al., Nat. Nanotech. 3, 46 (2008).

[15] C. G. Duan et al., Phys. Rev. Lett. 101, 137201 (2008).

[16] G. Catalana, Appl. Phys. Lett. 88, 102902 (2006).

[17] C. W. Nan et al , Phys. Rev. Lett. 94, 197203 (2005).

[18] G. Burns and B. A. Scott, Phys. Rev. Lett. 25, 1191 (1970); Phys. Rev. B 7, 3088 (1973); Phys. Rev. B 30, 7170 (1984).

[19] I. A. Akimov et al, Phys. Rev. Lett, 84, 4625 (2000); G. Blumberg et al. ibid, 78, 2461 (1997); J. F. Karpus et al. ibid, 93, 167205 (2004).

[20] H. C. He et al., Adv. Funct. Mater. 17, 1333 (2007); H. C. He et al., J. Appl. Phys. 103, 034103 (2008).

[21] C. Y. Deng et al., Acta Mater. 56, 405 (2008).

[22] T. F. O. Melo et al., Surf. Sci. 600, 3642 (2006).

[23] J. Frantti and V. Lantto, Phys. Rev. B 56, 221 (1997); A. G. Souza Filho et al., Phys. Rev. B 66, 132107 (2002); J. Frantti, et al., Ferroelectrics 266, 73 (2002).

[24] W.H. Xu et al. Appl. Phys. Lett.**79**(5), 411 (2001); J. Rouquette et al., Phys. Rev. B 73, 224118 (2006); J. R. Cheng et al., Appl. Phys. Lett. 88, 152906 (2006); M. Deluca et al., J. Eur. Ceram. Soc 26, 2337 (2006).

[25] W. H. Wang and X. Ren, J. Cryst. Growth 289, 605 (2006).

Table I. Frequencies of the mode vibrations in the PZT-CFO bilayer film in the presence of magnetic field, obtained from best numerical fitting by using Gaussian function to Raman spectra.

| | Mode (cm$^{-1}$) | Magnetic field (T) | | | |
|---|---|---|---|---|---|
| | | 0 | 0.08 | 0.16 | 0.28 |
| PZT layer | $A_1(TO_1)$ | 140 | 139 | 138 | 136.9 |
| | $E(TO_2)$ | 197.9 | 199.6 | 200.8 | 201.5 |
| | $B_1+E$ | 272.9 | 272.6 | 272.2 | 272.7 |
| CFO layer | $T_{2g}$ | 465.5 | 467.5 | 468.1 | 469.1 |
| | $A_{1g}$ | 686 | 688 | 688.7 | 690.1 |

**Figure captions**

FIG. 1 (color online). (a) Raman spectra of PZT-CFO bilayer film, and single-phase PZT and CFO films, without magnetic field. (b) Comparison of the soft-mode $A_1(TO_1)$ and $E(TO_2)$ phonon of PZT in the bilayer film and single PZT film, and of $A_{1g}$ mode of CFO in the bilayer film and single CFO film. Above (b) are structural units for the soft-mode $A_1(TO_1)$ and $E(TO_2)$ phonon of PZT. Spectra are shifted vertically for clarity.

FIG. 2 (color online). (a) Ferroelectric and (b) magnetic hysteresis loops of the PZT-CFO bi-layer and single phase films. The experimental electric field is an average voltage divided by the total thickness of the films. The appearance coercivity of the bilayer films is higher than that of the pure PZT film. The appearance polarization values of the bilayer film are bit lower than those for the pure PZT film due to the paraelectric CFO layer. The out-of-plane magnetization is normalized by the total volume of the films.

FIG. 3 (color). Raman spectra of (a) the PZT-CFO bilayer and (b) the PZT film in the presence of magnetic field. (c) Comparison of the changes of the soft-mode $A_1(TO_1)$ with magnetic field for the PZT-CFO bilayer and the PZT film. The lines in (c) are Gaussian fitting.

FIG. 4. Relative magnetic-field-induced changes of the soft-mode frequency (a) and the polarization (b). (a) and (b) correspond to the right- and left-hand sides of Eq. (1), respectively. The ME measurement was performed by also applying $H$ normal to the film plane with an ac magnetic field of 1.2 mT at 1 kHz superimposed as before (Ref. [21]).

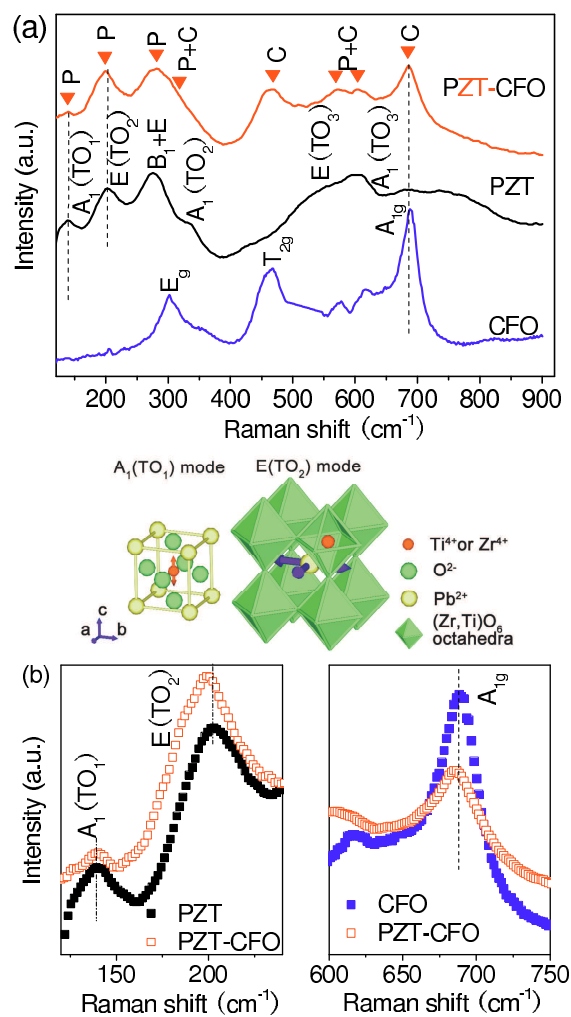

Figure 1

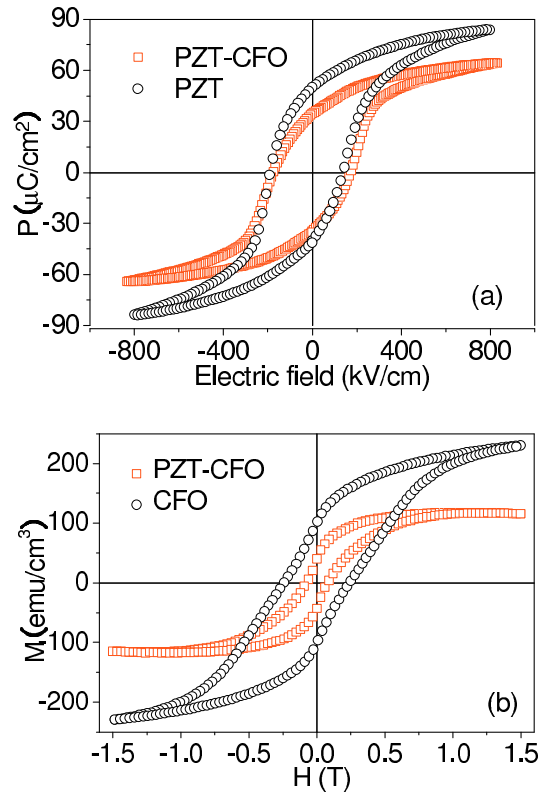

Figure 2

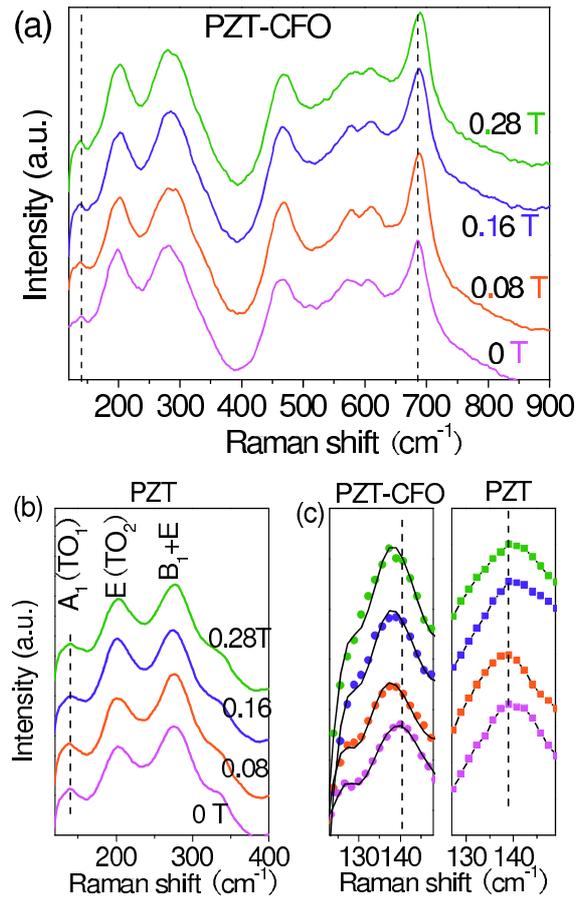

Figure 3

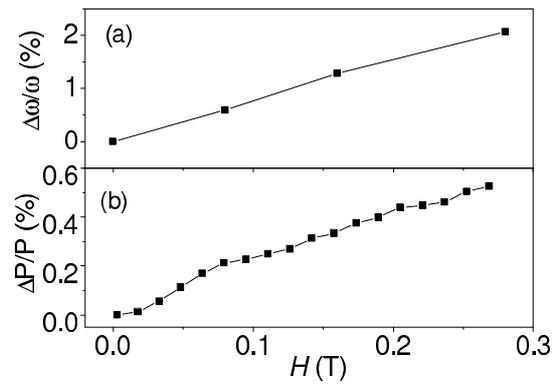

Figure 4